\documentstyle[aps,prl,floats,graphicx,preprint]{revtex}
\begin{document}
 \twocolumn[\hsize\textwidth\columnwidth\hsize\csname
@twocolumnfalse\endcsname

\title{The internal magnetic field in superconducting ferromagnets}
\author{Grigory I. Leviev\thanks{%
email: gileviev@vms.huji.ac.il},
Menachem I. Tsindlekht, Edouard B. Sonin, and Israel Felner}
\address{The Racah Institute of Physics, The Hebrew University of
Jerusalem,\\
91904 Jerusalem, Israel}

\date{\today} \maketitle

\begin{abstract}
We have measured the nonlinear response to the ac magnetic field
in the superconducting weak ferromagnet Ru-1222, at different
regimes of sample cooling which provides unambiguous evidence of
the interplay of the domain structure and the vorticity in the
superconducting state. This is {\em direct} proof of coexistence
of ferromagnetic and superconductive order parameters in
high-$T_c$ ruthenocuprates.

\end{abstract}
 ]

The problem of coexistence of superconductivity (SC) and
ferromagnetism (FM) has been studied for almost 50 years starting
from the theoretical work by Ginzburg \cite{G} (see also
\cite{BUL}). Coexistence of weak-ferromagnetism (W-FM) and SC was
discovered some time ago in RuSr$_2$R$_{2-x}$Ce$_x$Cu$_2$O$_{10}$
(R=Eu and Gd, Ru-1222) layered cuprate systems\cite{F}, and more
recently \cite{BER} in RuSr$_2$GdCu$_2$O$_8$ (Ru-1212). The SC
charge carriers originate from the CuO$_2$ planes and the W-FM  is
related to the Ru layers. In both systems, the magnetic order does
not vanish when SC sets in at $T_c$, and remains unchanged and
coexists with the SC state. The Ru-1222 materials (for R=Eu and
Gd) display a magnetic transition at $T_N= 125-180$ K and bulk SC
below $T_c$ = 25-50 K ($T_N >T_c$) depending on the oxygen
concentration and sample preparation. This discovery has launched
a new wave of investigations in this field \cite{exs}. The problem
is of general interest for condensed matter physics and is
relevant for many materials, in particular unconventional
superconductors (including some heavy fermions) with triplet
pairing \cite{HF}.

Despite a lot of work done in the past and recently, debates
concerning whether such coexistence is genuine are still
continuing. Evidence in favor of coexistence is mostly indirect
and refers to some peculiarities of the magnetization curve. One
of the most pronounced manifestations of SC-FM coexistence is the
spontaneous vortex phase (superconducting vortices induced by the
internal magnetic field from the FM magnetization). It explains
well the magnetization curve of these materials (see \cite{SF} and
references therein). However, this phase has not yet been observed
experimentally (visualized as the more common mixed state of type
II superconductors).

In the past the evidence of the FM-SC coexistence referred mostly
to the magnetic properties of the materials affected by the
presence of superconductivity. In this letter we present the first
experimental evidence of the effect of the ferromagnetic order
parameter, on the superconducting order parameter. The
ferromagnetic order parameter, namely the spontaneous
magnetization, is a source of an internal magnetic field inside a
sample even without an external magnetic field $H$. On the other
hand, the superconducting properties of type-II superconductors
depend strongly on whether the sample was cooled to the SC state
in zero magnetic field (ZFC) or in a finite magnetic field (FC).
Here a ``field'' is supposed to be an \textit{external} magnetic
field. We show here that these properties depend also on the {\em
internal} magnetic field during the cooling process. We exploited
the procedure, which we shall call the internal-field cooling
(IFC): The sample was cooled down to $T_{IFC}$ under an external
magnetic field $H_{IFC}$, ($T_{IFC} <T_N$). At $T_{IFC}$,
$H_{IFC}$ was turned off and further cool-down to $T=5$ K was done
at $H=0$. It appears that, by using the IFC procedure, the
properties of the SC state were different from those measured
after the regular ZFC process from temperatures above $T_N$. Thus,
in the SC state, the sample senses the internal magnetic field
evolved from the remanent magnetization, which was formed in the
normal ferromagnetic phase and then frozen at further cooling.

\begin{figure}
     \begin{center}
       \leavevmode
       \includegraphics[width=0.9\linewidth]{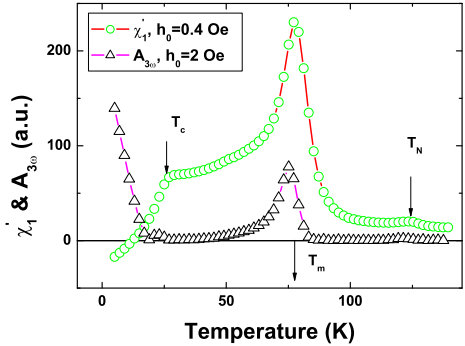}
       \bigskip
       \caption{Temperature dependencies of $\chi^{'}_{1}$ and $A_{3\omega}$}
     \label{f-1}
     \end{center}
     \end{figure}

We measured the nonlinear response to the ac magnetic field, which
is a sensitive probe of superconducting vorticity, as demonstrated
by numerous investigations in the past \cite{SSY,GOLD,GI}. Ceramic
sample of Gd$_{1.5}$Ce$_{0.5}$Ru$_{2}$Sr$_2$Cu$_2$O$_{10}$
(Ru-1222) with dimensions $8\times 2\times 2$ mm$^3$ was prepared
by a solid-state reaction as described in Ref. 2. In a nonlinear
medium, magnetization oscillations, induced by an ac magnetic
field $h(t)=h_0\sin\omega t$, may be expanded in a Fourier series:
\begin{equation}\label{eq1}
     M(t)=h_0\sum_{n >0} \chi^{'}_{n} \sin(n\omega t)-\chi^{''}_{n}
\cos(n\omega t)
\end{equation}
where $\chi^{'}_{n}$ and $\chi^{''}_{n}$ ($n = 1, 2,3...$) are the
in-phase and out-of-phase components of the harmonic
susceptibility. In all experiments described here we measured the
voltage drop induced in a pickup coil, which is proportional to
the time derivative of $M(t)$. Our home made experimental setup
was adapted to a commercial MPMS SQUID magnetometer. An ac field
$h(t)$ at a frequency of $\omega/2\pi = 1.5$ kHz and an amplitude
up to the $h_0= 3$ Oe was generated by a copper solenoid existing
inside the SQUID magnetometer. The temperature, dc magnetic field,
and amplitude dependencies of the fundamental and third harmonic
signals presented here have been measured by the two coils method
\cite{GOLD} . In the present letter the results for the first and
third harmonics will be discussed.

Figure \ref{f-1} shows the temperature dependencies of the
in-phase susceptibility $\chi^{'}_{1}$ and of the amplitude of the
third harmonic $A_{3\omega} \propto
h_0|\chi^{'}_{3}-i\chi^{''}_{3}|$, measured after the ZFC process
at $H=0$. The temperature dependence of $\chi^{'}_{1}$ is typical
for superconducting ferromagnets \cite{F}.  This  plot reveals
three transitions: (i) the paramagnetic-antiferromagnetic
transition at $T_N \approx 125$ K, (ii) the most pronounced
transition, which corresponds to the peak at $T_m \approx 78$ K,
and (iii) the transition into the SC state at $T_c \approx 28$ K.
The nature of the second transition, which is evident both in the
linear and the nonlinear response, is not yet completely clear and
is discussed elsewhere \cite{F,CAR}. Ambiguity is connected with
the magnetic phase between $T_m$ and $T_N$, which is characterized
by low coercivity. On the other hand, the $T_c<T<T_m$ temperature
region definitely corresponds to the weak ferromagnetic
phase \cite{F}.

\begin{figure}
      \begin{center}
        \leavevmode
        \includegraphics[width=0.9\linewidth]{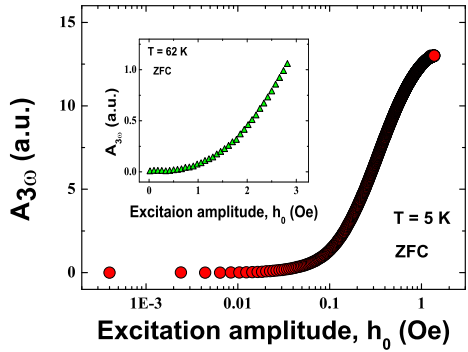}
        \bigskip
        \caption{Amplitude dependencies of $A_{3\omega}$ at $T = 5$K.
Inset: amplitude
        dependence of  $A_{3\omega}$ in magnetic phase at $T = 62$ K.}
      \label{ampl}
      \end{center}
      \end{figure}

The third harmonic behavior is different for $T>T_c$ and $T<T_c$.
For $T>T_c$ the behavior is typical for ferromagnetic materials
and was known already from Rayleigh's investigation on iron
\cite{LR}. The third harmonic response demonstrates a quadratic
dependence on $h_0$ (inset in Fig. \ref{ampl}), which directly
derived from the oscillatory motion of the domain walls \cite{B}.
This signal should decrease at low temperatures and it becomes
unobservable under our experimental conditions. For $T<T_c$ the
third harmonic grows very fast with temperature decreasing (Fig.
\ref{f-1}), and its dependence on the ac field amplitude (Fig.
\ref{ampl}) is different from that at $T>T_c$, as evident from the
saturation for the nonlinear response at high amplitude of
excitation, instead of a quadratic growth. The growth of the
nonlinear response in the superconducting materials was revealed
in numerous previous experimental investigations
\cite{SSY,GOLD,GI}. Various mechanisms were suggested for this
nonlinear response based on the critical state model \cite{SSY}
and the presence of weak links \cite{GI}. In particular, the
response shown in Fig. \ref{ampl} is well described by the
Josephson-media model. We do not have to discuss these models,
since  all of them relate the response to the penetration of the
magnetic flux (vortices) into the sample, and only this fact is
essential for the present investigation. Thus it seems reasonable
that the $A_{3\omega}$ at $T<T_c$ is an effective probe of the
superconducting vorticity.

\begin{figure}
     \begin{center}
       \leavevmode
       \includegraphics[width=0.9\linewidth]{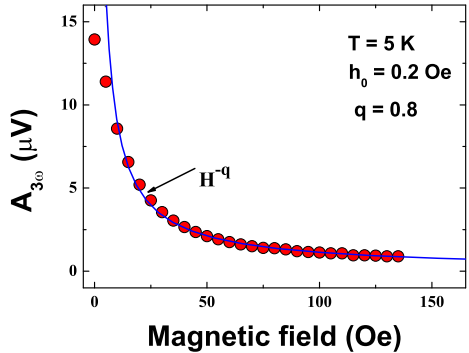}
       \bigskip
       \caption{Magnetic field dependence of $A_{3\omega}$ at $T=5$ K after
ZFC}
     \label{f-2}
     \end{center}
     \end{figure}

  Figure \ref{f-2} demonstrates ZFC dependence of
$A_{3\omega}$ on the external magnetic field. One can see that
$A_{3\omega}$ decreases with the magnetic field. At high magnetic
fields $A_{3\omega}$ is a power function of the $H$:
$A_{3\omega}\propto H^{-q}$, with $q \approx 0.8$. Suppression of
the $A_{3\omega}$ by the magnetic field applied after ZFC was
observed in the previous nonlinear studies and agrees with all
suggested models of the nonlinear response. The nonlinearity under
discussion is connected with a nonhomogeneous distribution of the
magnetic flux, which penetrated into the sample, and the magnetic
flux distribution becomes more and more uniform, when the vortex
density increases. On the other hand, in the Meissner state the
nonlinear response must be quite weak, and the magnetic field
dependence of $A_{3\omega}$ should have a peak at some $H$, as was
observed in some materials \cite{SSY}. But in ceramics with
numerous weak links, such as our material, this field can be
extremely small, and the peak is not observable. Moreover, we deal
with the superconducting ferromagnets, where the spontaneous
vortex phase can replace the Meissner state at $H=0$. Altogether
this explains why we observe the maximum value of $A_{3\omega}$ at
$H=0$.

\begin{figure}
     \begin{center}
       \leavevmode
       \includegraphics[width=0.9\linewidth]{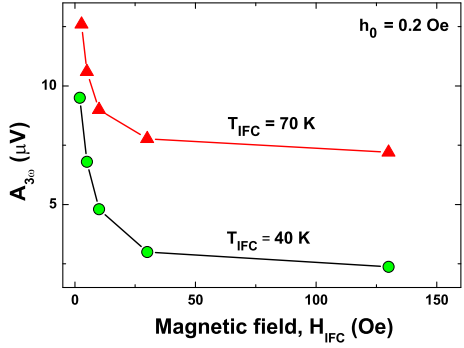}
       \bigskip
       \caption{$A_{3\omega}(T=5 K)$ as a function of $H_{IFC}$ for $T_{IFC}
= 40$ K and 70 K}
     \label{f-2a}
     \end{center}
     \end{figure}

Now let us consider the experimental results in the IFC process.
After turning off the magnetic field $H_{IFC}$ at temperature
$T_{IFC}$, the sample was cooled in $H=0$ down to $T=5$ K and the
signal of the third harmonic at $T=5$ K was measured. Figure
\ref{f-2a} shows $A_{3\omega}(H_{IFC})$ dependence for $T_{IFC} =
40$ K and $70$ K. It is evident that the field $H_{IFC}$
suppresses the $A_{3\omega}$ signal similarly to the external
field after ZFC Fig. \ref{f-2} even though $H_{IFC}$ was turned
off {\textit{before}} the onset of superconductivity. Turning off
$H_{IFC}$ at $T=40$ K affects $A_{3\omega}$ more strongly than for
$T = 70$ K due to larger remanent magnetization at $T=40$ K. This
behavior is typical for the FM materials \cite{B}.

\begin{figure}
     \begin{center}
       \leavevmode
       \includegraphics[width=0.9\linewidth]{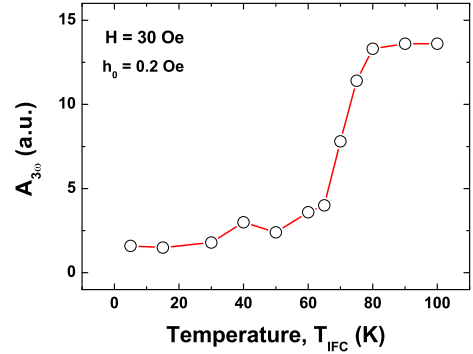}
       \bigskip
       \caption{Amplitude of the third harmonic $A_{3\omega}$ at $H = 0$ and
$T = 5 $ K vs $T_{IFC}$}
     \label{f-2c}
     \end{center}
     \end{figure}

\begin{figure}
     \begin{center}
       \leavevmode
       \includegraphics[width=0.9\linewidth]{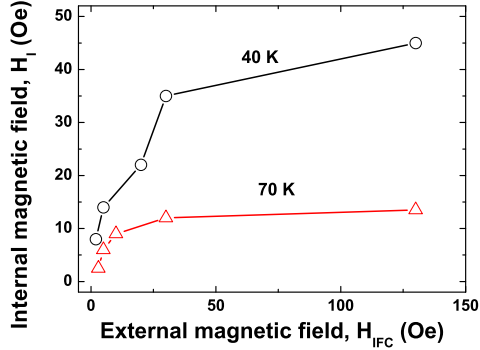}
       \bigskip
       \caption{Internal magnetic field vs $H_{IFC}$ for $T_{IFC} = 40$ K and
70 K.}
     \label{f-3}
     \end{center}
     \end{figure}

Figure \ref{f-2c} presents the signal of the third harmonic
$A_{3\omega}(T = 5 K)$ as a function of $T_{IFC}$ after cooling in
$H_{IFC} = 30$ Oe. The signal of the $A_{3\omega}(T = 5 K)$
decreases for $T_{IFC} < T_m$. This demonstrates that the
suppression of the third harmonic response by the internal
magnetic field takes place only if the field cooling continues
down to the weakly ferromagnetic phase with essential coercivity.
It is known \cite{SF} that in idealized single-domain
superconducting ferromagnets the internal magnetic field from the
spontaneous magnetization $4\pi \vec M$ has the same effect on the
phase diagram, i.e., on the magnetic flux penetrating into the
sample, as the external field. This can be generalized in the more
realistic case of a multi-domain sample with nonzero average
internal field $4\pi \langle \vec M\rangle$. On the basis of this
argument we can use plot of $A_{3\omega}(H)$ (Fig. \ref{f-2}) as a
calibration curve to estimate the magnitude of the frozen internal
magnetic field ($H_I$). Namely, we take the value of $A_{3\omega}$
from the plot in Fig. \ref{f-2a}, find the value of $H$, which
corresponds to this value of $A_{3\omega}$ in  Fig \ref{f-2}, and
assume that this value of $H$ gives a reasonable estimation of
$H_I$. Figure \ref{f-3} presents the dependence of $H_I$ on
$H_{IFC}$.

The internal magnetic field arises from the frozen remanent
magnetization $4\pi\langle \vec M\rangle$ after field cooling down
to $T_{IFC}$. We have compared obtained in Fig. \ref{f-3} with
direct dc remanent magnetization measured in our previous studies
\cite{RM}. It appears that there is a reasonable agreement (with
an accuracy of $\pm20\%$) between the two methods, and confirms
our scenario.

The phenomenon revealed in our experiment is possible if the
domain structure formed in the ferromagnetic phase can be frozen
down to the superconducting state. On the other hand, as was noted
in the pioneering paper by Ginzburg \cite{G} and confirmed by the
detailed analysis in Ref. \cite{SONIN}, superconductivity should
strongly affect the equilibrium domain structure: Its period
should grow, and in equilibrium any sample in the Meissner state
is a single domain. But in our case we deal with a non-equilibrium
domain structure, which is a metastable state due to coercivity.

The presence of the frozen internal field in the superconducting
phase clearly demonstrates that the sample is in the mixed state
with many vortices. One cannot call this state the spontaneous
vortex phase because the latter refers to the {\em equilibrium}
state, but we deal with a metastable state. We have analyzed here
the nonlinear response, which is sensitive to the average internal
field $4\pi\langle \vec M\rangle$. The absolute value of the
average magnetization $\langle \vec M\rangle$ is less than the
saturation magnetization $M$, which can determine the vortex
density in a single-domain sample \cite{SF}. However, the
saturation magnetization may create vortices inside domains. Since
$\vec M$ changes its direction from domain to domain, we obtain
the vortex tangle, which does not contribute to the average
internal field $\sim 4\pi\langle \vec M\rangle$, studied here.
This vortex tangle is expected to exist even after the ZFC process
and contributes to the initial value of the third harmonic, which
was detected without external or internal magnetic field. These
arguments illustrate that the vorticity (magnetic flux)
distribution in a real (especially ceramic) superconducting
ferromagnet can be very complicated. Genuinely zero field cooling
is practically impossible: if one cools a sample in zero external
field, one cannot avoid internal magnetic fields from the
spontaneous magnetization, even if these fields vanish on average
but  still remain inside the domains. A more detailed analysis of
the magnetic-flux distribution  would become possible  if further
investigations provided more information on the structure of the
material: sizes of grains and domains, data on crystal anisotropy
etc.

In summary, our measurements of the nonlinear response
unambiguously demonstrate the coexistence of the superconducting
and ferromagnetic order parameter in Ru-1222 samples below the
superconducting critical temperature. Coexistence is manifested by
the clear effect  on the domain structure frozen from normal FM
phase on superconducting properties. We tend to believe that the
effect revealed in Ru-1222 is a general and can be observed in
other materials with FM-SC coexistence.

The work was supported by the Klatchky Foundation and by a grant
from the Israel Academy of Sciences and Humanities.


\begin{references}
\bibitem{G} V. L. Ginzburg,  Zh. Eksp. Teor. Fiz. {\bf 31}, 202 (1956)
[Sov. Phys.-JETP, {\bf 4}, 153 (1957)].

\bibitem{BUL} L.N. Bulaevskii, A.I. Buzdin, M.L. Kulic, and S.V.
Panyukov, Adv. Phys. {\bf 34}, 176 (1985); "Superconductivity in
Ternary Compounds" Vol II, edited by M.B. Maple and  O.  Fisher
(Springer-Verlag, Berlin, 1982).


\bibitem{F}I. Felner, U. Asaf, Y. Levi, and O. Millo, Phys. Rev.
B {\bf 55}, R3374 (1997).


\bibitem{BER} C. Bernhard, J.L. Tallon, Ch. Niedermayer, Th.
Blasius, A. Golnik, B. Brucher, R.K. Kremer, D.R. Noakes, C.E.
Stronach, and E.J. Ansaldo, Phys. Rev. B {\bf 59}, 14099 (1999).


\bibitem{exs} Y.Y. Xue, B. Lorenz, D.H. Cao, and C.W. Chu, Phys.
Rev. B, {\bf 67}, 184507 (2003); W.E. Pickett, R. Weht, and A.B.
Shick, Phys. Rev. Lett. {\bf 83}, 3713 (1999).



\bibitem{HF}M. Sigrist, Progr. Theor Phys.
{\bf 99}, 899 (1998); T. Nishioka, G. Motoyama, S. Nakamura, H.
Kadoya, and N.K. Sato, Phys. Rev. Lett. {\bf 88}, 237203 (2002);
A.P. Mackenzie and Y. Maeno, Rev. Mod. Phys. {\bf 75}, 657 (2003).



\bibitem{SF} E. B. Sonin and I. Felner, Phys. Rev. B {\bf 57},14000
(1998).

\bibitem{SSY} S. Shatz, A. Shaulov, and Y. Yeshurun, Phys. Rev. B
{\bf 48}, 13 871 (1993).

\bibitem{GOLD} T. Ishida and R.B. Goldfarb, Phys. Rev. B {\bf 41},
8937 (1990).

\bibitem{GI}  G.I. Leviev, A. Pikovsky, and D.W. Cooke, Supercond.
Sci. Technol. {\bf 5}, 679 (1992).



\bibitem{CAR} C.A. Cardoso, F.M. Araujo-Moreira, V.P.S. Awana,
E. Takayama-Muromachi, O.F. de Lima, H. Yamauchi, and M.
Karppinen, Phys. Rev. B {\bf 67}, 020407(R) (2003).




\bibitem{LR}   Rayleigh, Lord, Phil. Mag. {\bf 23}, 225 (1887).
\bibitem{B} R.M. Bozorth, {\it Ferromagnetism}, D. Van Nostrand
Company, Inc., NY (1951).





\bibitem{SONIN} E.B. Sonin, Phys. Rev. B {\bf 66}, 100504
(2002).


\bibitem{RM} I. Felner and E. Galstyan, Physica C (2003), in
press.


\end{references}
\end{document}